\documentclass{article}

\usepackage{arxiv}
\usepackage{tabularx}
\usepackage[ruled,linesnumbered,commentsnumbered]{algorithm2e}
\usepackage[T1]{fontenc}
\usepackage[utf8]{inputenc}
\usepackage{graphicx}
\usepackage[T1]{fontenc}
\usepackage{mathptmx}
\usepackage{hyperref}
\usepackage{tablefootnote}
\usepackage[nottoc]{tocbibind}
\usepackage{caption}
\usepackage{subcaption} 
\usepackage{footnote}
\usepackage{amsmath}
\usepackage{amssymb}
\usepackage{mathrsfs}
\usepackage{array}
\usepackage{dcolumn}
\usepackage{bm}
\usepackage[utf8]{inputenc} 
\usepackage[T1]{fontenc}    
\usepackage{hyperref}       
\usepackage{url}            
\usepackage{booktabs}       
\usepackage{amsfonts}       
\usepackage{nicefrac}       
\usepackage{microtype}      
\usepackage{lipsum}
\usepackage{framed}

\usepackage[utf8]{inputenc}
\title{Measuring Causality: The Science of Cause and Effect}

\author{
Aditi Kathpalia$^{1,2}$, Nithin Nagaraj$^{1}$\\
$^{1}$Consciousness Studies Programme,\\ National Institute of Advanced Studies, Indian Institute of Science Campus, Bengaluru 560012, India. \\
$^{2}$Manipal Academy of Higher Education, Manipal, Karnataka 576104, India. \\
\texttt{kathpaliaaditi@nias.res.in, nithin@nias.res.in  } \\
}
\begin{document}
\maketitle

\begin{abstract}
Determining and measuring cause-effect relationships is fundamental to most scientific studies of natural phenomena. The notion of causation is distinctly different from correlation which only looks at association of trends or patterns in measurements. In this article, we review different notions of causality and focus especially on measuring causality from time series data. Causality testing finds numerous applications in diverse disciplines such as neuroscience, econometrics, climatology, physics and artificial intelligence.
\end{abstract}

\keywords{Causality \and correlation \and ladder of causation \and Granger causality \and model-based causality}

\section{\label{Sec_Intro} Introduction}

Most studies in natural as well as social sciences are centred around the theme of determining cause-effect relationships between processes or events. Such studies are being conducted from the early 20th century onwards. While some studies are observational, others involve experiments to understand the nature of dependencies. Examples of observational studies involve, studying the particle size and fertility of soil, availability of water, diseases or pests in a particular place in order to study their effect on crop yield; or observing the death rates of smoking vs non-smoking people to determine its influence on mortality. On the other hand, an example of experimental study would be studying a diseased group of people who are being administered medication to check its efficacy against a control group of people being administered a similar dose of placebo drug. 

\subsection*{Three types of statistical causality}
Cox and Wermuth have given three notions (levels) of statistical causality based on existing approaches for estimating causality~\cite{Cox_Wermuth_2004}. The zero-level view of causality is basically a statistical association, i.e. non-independence with the cause happening before the effect. This association cannot be done away with by conditioning on alternative allowable features. For example, when looking at the causal influence that greenhouse gases in the atmosphere have on increasing temperature of earth's surface, other features such as solar output which are also potential causes of the effect in question need to be conditioned. Only then can greenhouse gases be said to have an effect on earth's temperature. In mathematical terms, it is a dependence based on a multiple-regression like analysis that cannot be explained by other appropriate explanatory variables. This type was studied by Good (1961,1962)~\cite{Good1, Good2} and by Suppes (1970)~\cite{Suppes}. In a time-series context, it was formalized as Wiener-Granger causality by Granger (1969)~\cite{Granger} and later, formulated in a more general context by Schweder (1970)~\cite{Schweder} and by Aalen (1987)~\cite{Aalen}.

In the first-level view of causality, the aim is to compare the outcomes arising under different interventions, given two or more (possible) interventions in a system. For example, take the case of two medical interventions, $D_1$ and $D_0$ -- a treatment drug and a control respectively, only one of which can be given to a particular patient. The outcome observed with $D_1$ use is compared with the outcome that would have been observed on that patient had $D_0$ been used, other things being equal. If there is evidence that use of $D_1$ instead of $D_0$ causes a change in outcome, then it can be said that $D_1$ \emph{causes} that \emph{change}. The key principles of such kind of experimental design for \emph{randomized control trials} were developed mainly at Rothamsted (Fisher, 1926, 1935; Yates, 1938, 1951)~\cite{Fisher1, Fisher2}. This way of inferring causation may have an objective of decision-making or may require the conduction of a controlled experiment, although that is not always the case. For
example, when trying to check if an anomalous gene is the cause of a particular disease, the intervention as between the abnormal and normal version of the gene is hypothetical (since explicit intervention is not possible) and also no immediate decision-making process is generally involved. Rubin (1974)~\cite{Rubin} adapted the notions of causality to observational studies using a representation similar to Fisher's. The definition of causality in the above discussed first-level view is explicitly comparative and has been the most widely used in scientific studies.

Suppose that preliminary analysis in a scientific context has established a pattern of associations/ dependencies or have provided good amount of evidence of first- or zero-level causality. Second-level causality is used for explaining \emph{how these dependencies arose} or \emph{what underlying generating process were involved for the causal relationships observed}. On several occasions, this will require incorporating information from previous studies in the field or by doing laboratory experiments. Attempts in this regard started with graphical representations of causal path diagrams by Sewall Wright (Wright, 1921, 1934)~\cite{Wright1, Wright2} and was later promoted by Cochran (1965)~\cite{Cochran}. Currently, Non Parametric Structural Equations Models (NPSEMs) (Pearl, 2000)~\cite{Pearl1} which provide a very general data-generating mechanism
suitable for encoding causation, dominate the field.

Each of the above types for determining causality have their own pros and cons and their use depends on the motive and the nature of the study. While first-level causal estimation that mostly involves randomization experiments may make the conclusions of the study more secure, but fails to reveal the biological, or psychological, or physical processes working behind
the effect observed. On the other hand, zero-level causality suffers from the criticism that their is no \emph{intervention} involved to observe the causal effect of \emph{doing} something on the system. The second-level of causality requires field knowledge and cannot be solely data driven.

While it is useful to know all these notions of causality, for the rest of this article, we will mostly deal with causality as estimated from collected time series measurements where it is not possible to intervene on the experimental setup. 

%
\section{Correlation and Causation}

We have often heard the saying `Correlation does not imply Causation'. But even to this date, there are several scientific studies which make erroneous conclusions regarding a variable being a cause of another, merely on the basis of observed correlation value. Thus it becomes necessary to clarify the meaning and use of these two terms.

\begin{figure}[!t]
\vskip -12pt
\centering
\includegraphics[width=0.9\textwidth,trim={0 0 0 0},clip]{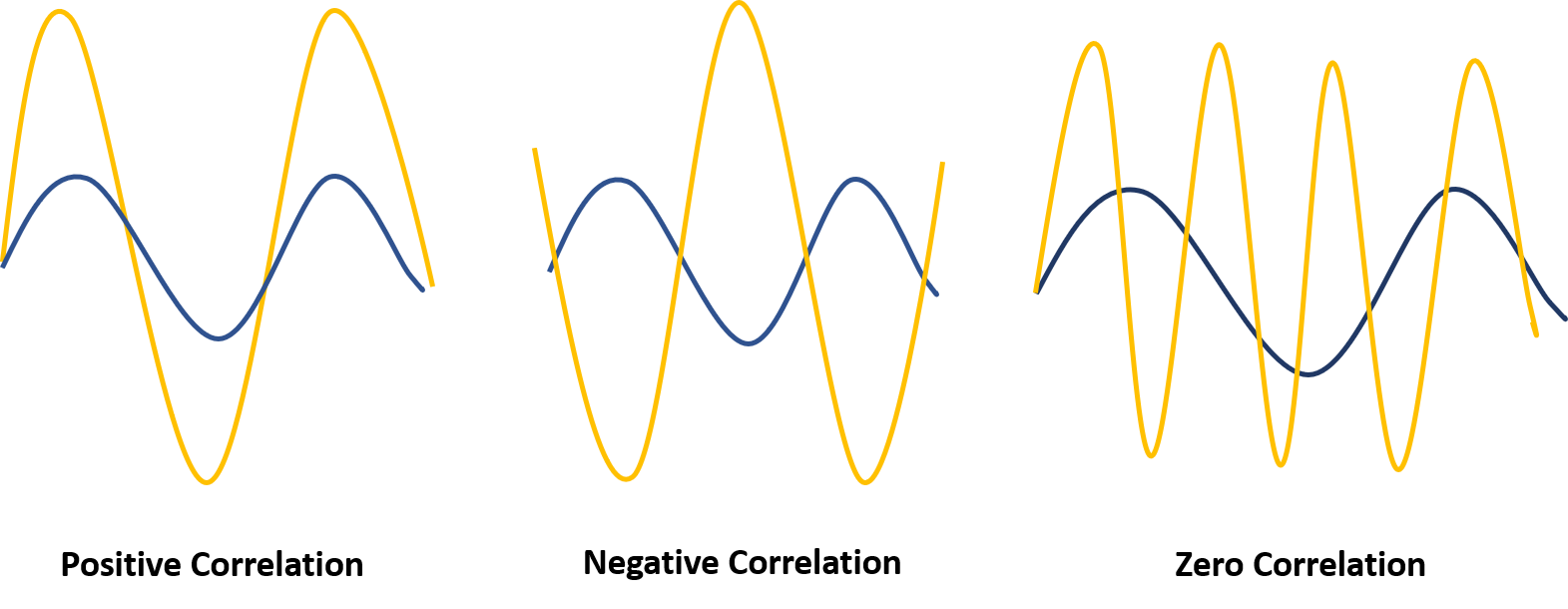}
\caption{Positive, negative and zero correlation.}\label{fig_corr_trends}
\end{figure}

Correlation is a statistical technique which tells how strongly are a pair of variables linearly related and change together. It does not tell us the `why' and `how' behind the relationship but it just tells that a mathematical relationship exists. For example, Pearson's correlation coefficient for a pair of random variables ($X,Y$) is given as:

\begin{equation}
    \rho_{X,Y}=\frac{\mathbb{E}[(X-\mu_X)(Y-\mu_Y)]}{\sigma_X\sigma_Y},
\end{equation}
where, the numerator is the covariance of variables $X,Y$ and $\sigma_X, \sigma_Y$ are the standard deviations of $X$ and $Y$ respectively. $\mathbb{E}$ is the expectation and $\mu_X, \mu_Y$ are the means of $X$ and $Y$ respectively. Note that:  $-1 \leq\rho_{X,Y} \leq +1$ and is always symmetric $\rho_{X,Y} = \rho_{Y,X}$. The closer the magnitude is to $1$, the stronger is the relationship between the variables. Figure~\ref{fig_corr_trends} illustrates two signals with positive, negative and zero correlation. An example of positive correlation would be between temperature in a region and sale of coolers -- as temperature increases (decreases), sale of coolers also increases (decreases). However, as temperature increases (decreases), the sale of heaters decreases (increases), indicating negative correlation. An example of zero correlation would be between the amount of tea consumed by an individual and his/her level of intelligence. %

In contrast, causation indicates that one event is a result of the occurrence of another event.  A variable $X$ can be said to be a \emph{cause} of another variable $Y$, ``if it makes a difference to $Y$ and the difference $X$ makes must be a difference from what would have happened without it". This definition is adapted from the definition of a `cause' given by philosopher David Lewis~\cite{Pearl2}. As discussed in the previous section, there are several means for estimating causality. Unlike correlation, causation is asymmetric.

Interestingly, for conventional statistics, causation was a non-scientific concept and as per the ideas prevalent in the late 19th and early 20th century, all analysis could be reduced to correlation. Since correlation got rigorously mathematically defined first (when scientist Galton was in search of a tool for causation) and causation seemed to be only a limited category of correlation, the latter became the central tool. Moreover, the pioneers of statistics such as Pearson felt that causation is only a matter of re-occurrence of certain sequences and science can in no way demonstrate any inherent necessity in a sequence of events nor prove with certainty that the sequence must be repeated~\cite{Pearl2}.

However, later on since most studies were in search of causal inferences and agents for their experimental/observational data and were at the same time using the famous statistical tool of correlation, they ended up incorrectly deducing the existence of causation based on results from correlation measures. Of the several infamous studies, an example is of the 2012 paper published in the \emph{New England Journal of Medicine} which claims that chocolate consumption can lead to enhanced cognitive function. The basis for this conclusion was that the number of Nobel Prize laureates in each country was strongly correlated with the per capita consumption of chocolate in that country. One error that the authors of the paper made was deducing individual level conclusions (regarding enhancement of cognitive level) based on group level (country) data. There was no data on how much chocolate Nobel laureates consumed. It is possible that the correlation between the two variables arose because of a common factor -- the prosperity of the country which affected both the access to chocolate as well as availability of higher education in the country.

\begin{figure}[!t]
\centering
\includegraphics[width=7.5 cm,trim={0 0 0 0},clip]{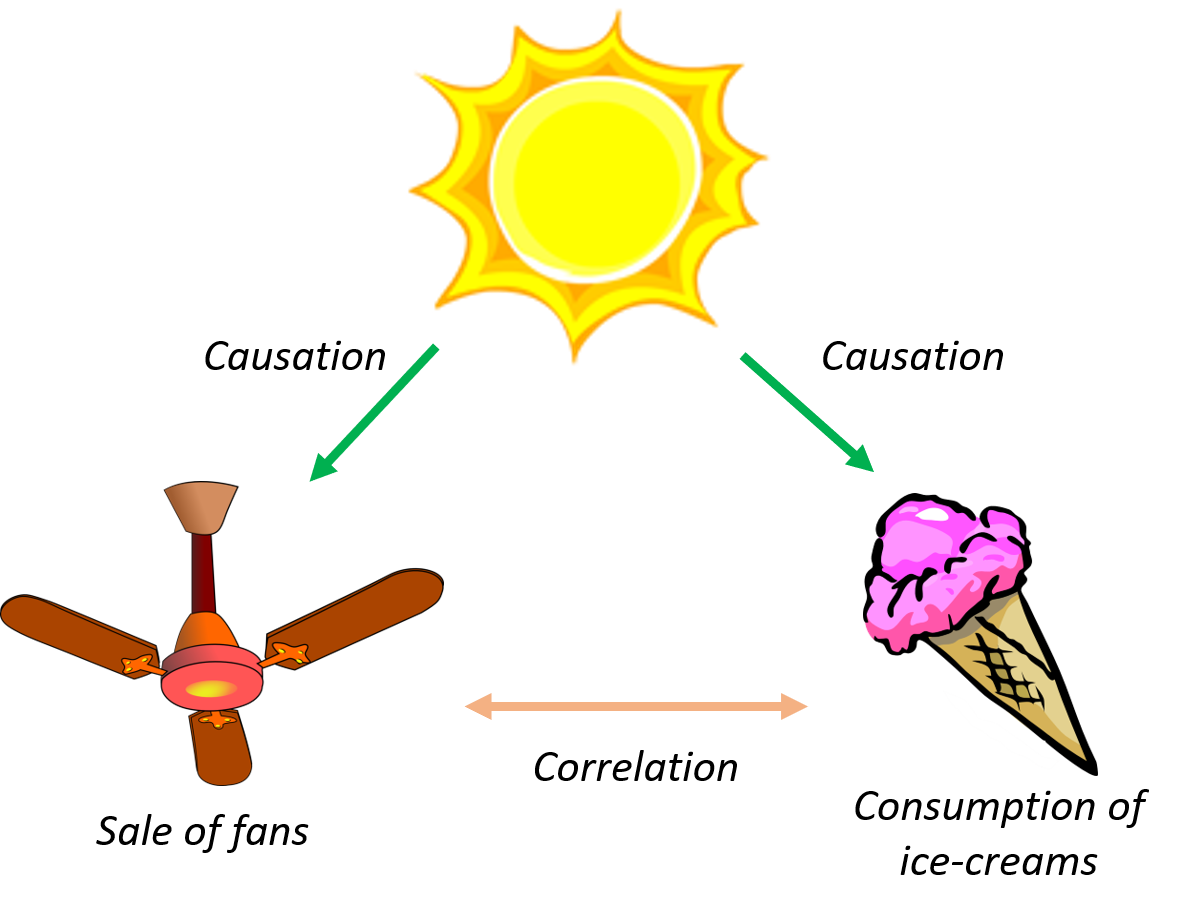}
\caption{High correlation between `sale of fans' and `consumption of ice-creams' as a result of a \emph{confounding variable}, `temperature in a region'.}\label{fig_confounding}
\end{figure}

There are several cases in everyday life where we can observe that correlation between two variables increases because of a common cause variable influencing the observed variables. This common cause variable is referred to as the \emph{confounding variable} which results in a spurious association between the two variables. Figure~\ref{fig_confounding} shows the example of the confounding variable `temperature in a region' influencing the observed variables `sale of fans' and `consumption of ice-creams', resulting in a high correlation between the latter two variables.

\section{The Ladder of Causation}

Judea Pearl, in his latest book, `The Book of Why', gives three levels for a causal learner~\cite{Pearl2}. His work on machine learning convinced him that for machines to learn to make decisions like humans, they cannot continue to make associations based on data alone but needed to have causal reasoning analogous to the human mind. In the `Ladder of Causation' that he proposes, the three levels are $-$ 1. Association, 2. Intervention and 3. Counterfactuals, when arranged from the lower rung to the higher rung.

Association involves observing regularities in the data to associate a past event with a future one. Animals learn in this way, for example, this is what a cat does when it observes a mouse and tries to predict where it will be a moment later. Pearl argues that machine learning algorithms even till today operate in this mode of `association'. Correlation based measures such as those discussed in Section~\ref{Sec_Intro} under zero-level view of causality, work based on association. Intervention, at a higher level than association, involves actively changing what is there and then observing its effect. For example, when we take a paracetamol to cure our fever, it is an act of intervention on the drug level in our body to affect the fever. Randomized control trials as well as model-based causality measures (which aim to find the underlying generating mechanism) fall in this category. These have been discussed in Section~\ref{Sec_Intro} as the first and second levels of causality. While model based measures do not directly intervene, they invert the assumed model to obtain its various parameters based on the available data. The complete model can then be helpful to intervene, such as to make predictions about situations for which data is unavailable. 

\begin{figure}[!b]
\centering
\includegraphics[width=0.8\textwidth,trim={0 0 0 0},clip]{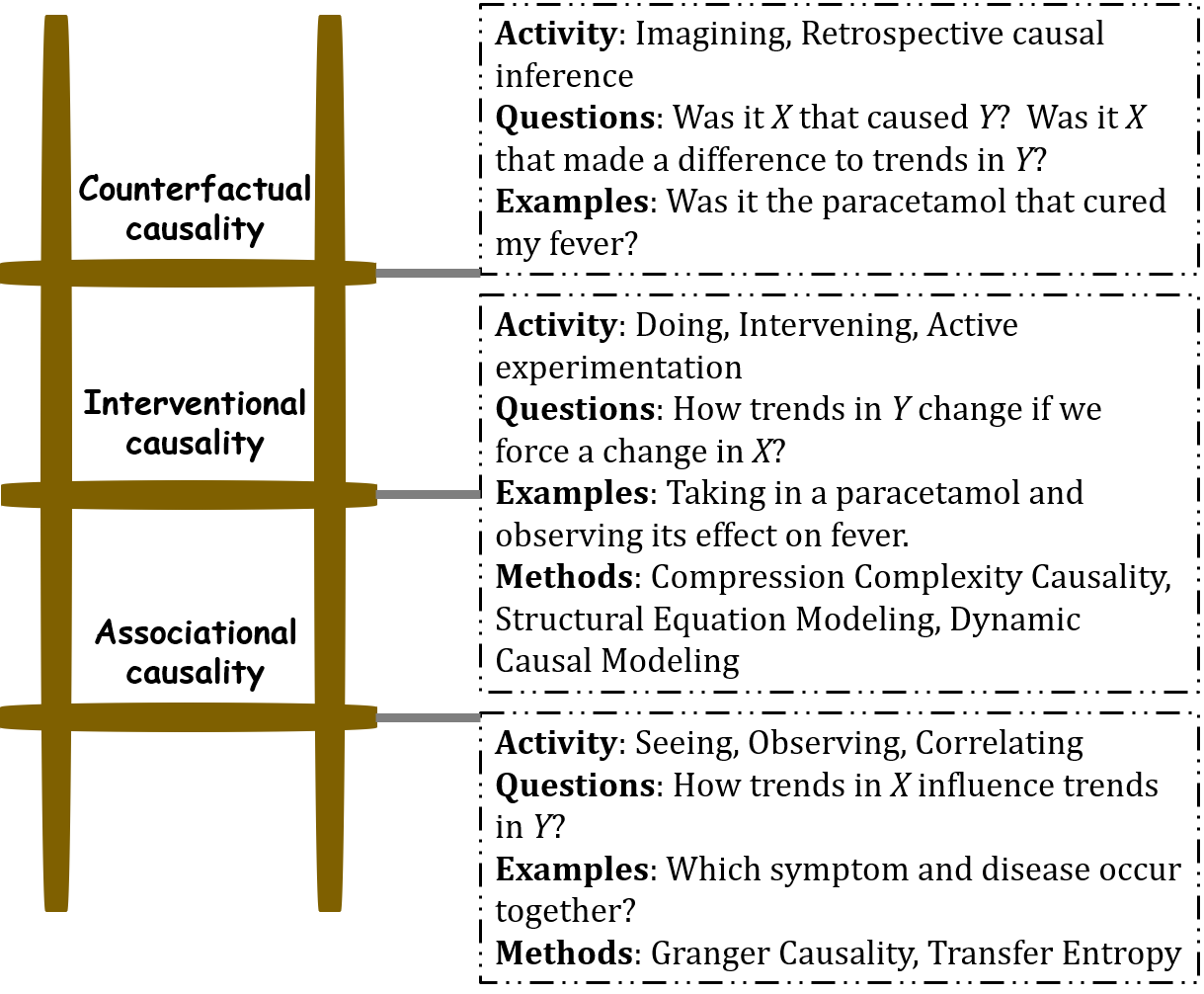}
\caption{The Ladder of Causation, adapted from~\cite{Pearl2}, for time series analysis.}\label{fig_ladder}
\end{figure}

The highest rung on the Ladder of Causation is that of Counterfactuals. This involves imagination. No experiment can actually change history (since time travel is not practical), but if I take paracetamol when I have fever and after a few hours I ask `was it the paracetamol that cured my fever?', then I am exercising the power of my imagination to infer the cause of my fever being cured. To date, there is no computational method to establish causality by such counterfactual reasoning.

We adapt Judea Pearl's ladder of causation to classify the methods of causality testing that are exclusively for time series data. Figure~\ref{fig_ladder} depicts the ladder, asking relevant `Questions' for causation from a time series perspective, giving `Examples' from everyday life and showing the time series analysis `Methods' that fall in each category. The methods are discussed in the following sections.

\section{Data-driven causality measures}

In the present day scenario, data is readily available and typically in large quantity. Also, to infer certain kinds of cause-effect relationships, it may be difficult or impossible to conduct intervention experiments. Thus, an increasing number of studies are now using data-driven measures of causality testing. While model-based causality measures would give more information about the underlying mechanism, when field knowledge is not adequately available, it may not be feasible to design such models. In such scenarios as well, model-free, data driven measures are useful. These are being employed in fields such as neuroscience, climatology, econometrics, physics and engineering (see Introduction of~\cite{Kathpalia}).

Several methods of causality which use time series data have been developed. One of the earliest and popular methods in this regard is Granger Causality (GC)~\cite{Granger}. Other methods that were proposed later include Transfer Entropy (TE)~\cite{Schreiber} and Compression-Complexity Causality (CCC)~\cite{Kathpalia}. All these methods are based on Wiener's idea~\cite{Wiener}, which defines a simple and elegant way to estimate causality  from time series data. According to Wiener, ``if a time series $X$ \emph{causes} a time series $Y$, then past values of $X$ should contain information that help predict $Y$ above and beyond the information contained in past values of $Y$ alone". Wiener's approach to causation and the idea behind different methods based on it is given in Box 1.
%
%
 \begin{framed}
\begin{center}
\textbf{Box 1} \\
\vspace{0.1in}
{\includegraphics[width=8cm]{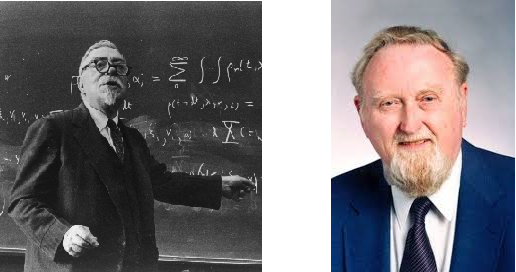}}\\
Norbert Wiener (1894-1964) (left) and Clive W.J. Granger (1934-2009) (right) -- pioneers in the field of time series based causality estimation. Granger was awarded the Nobel Memorial Prize in Economic Sciences in 2003 for his work on methods for analyzing economic time series with common trends.
\end{center}
\subsection*{Wiener's idea:}
According to Wiener, ``if a time series $X$ \emph{causes} a time series $Y$, then past values of $X$ should contain information that help predict $Y$ above and beyond the information contained in past values of $Y$ alone"~\cite{Wiener}. 

\begin{center}
\label{fig_wiener}
\vskip -12pt
\centering

\includegraphics[width=5.5cm,trim={0 0 0 0},clip]{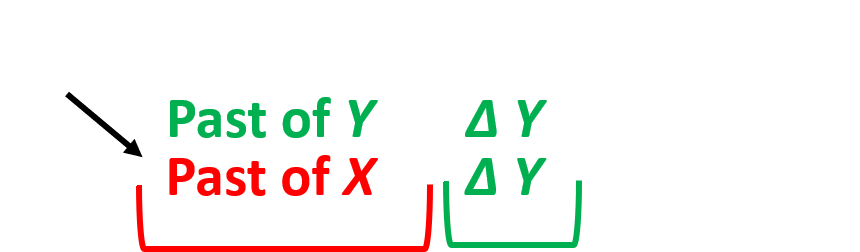}
\end{center}
Several methods are based on this approach and the idea behind each one of them is stated below. If, with the inclusion of past of $X$ –-

\begin{itemize}
    \item prediction power of $Y$ $\uparrow$, then there is a non-zero \textbf{Granger Causality} from $X$ to $Y$.
    \item uncertainty of $Y$ $\downarrow$, then there is a non-zero \textbf{Transfer Entropy} from $X$ to $Y$.
    \item dynamical complexity of $Y$ $\downarrow/ \uparrow$, then there is a non-zero \textbf{Compression-Complexity Causality} from $X$ to $Y$.
    \end{itemize}
\end{framed}

Other model-free methods for causality testing from time series data include Convergent Cross Mapping~\cite{Sugihara}, Topological Causality~\cite{Harnack} etc. These measures capture causality based on the topological properties of dynamical systems. We discuss a few important methods below.

\subsection{Granger Causality (GC)}
Granger causality is a statistical concept of causality that is based on prediction. This was the first method that was directly based on Wiener's approach and hence it is often referred to as Wiener-Granger Causality~\cite{Wiener}. To check if a process $X$ \emph{Granger causes} another process $Y$, two separate autoregressive processes of $Y$ are modeled for consideration --
\begin{equation}
    Y(t)=\sum_{\tau=1}^{\infty}(a_\tau Y(t-\tau))+\sum_{\tau=1}^{\infty}(c_\tau X(t-\tau))+\varepsilon_c,
    \label{eq_granger_causal_model}
\end{equation}
    
\begin{equation}
Y(t)=\sum_{\tau=1}^{\infty}(b_\tau Y(t-\tau))+\varepsilon,
 \label{eq_granger_noncausal_model}
\end{equation}

where $t$ denotes any time instance,  $a_\tau, b_\tau, c_\tau$ are coefficients at a time lag of $\tau$ and  $\epsilon_c, \epsilon$ are error terms in the two models. Assuming that $X$ and $Y$ are covariance stationary\footnote{A process is said to be covariance (or weak-sense) stationary if its mean does not change with time and the covariance between any two terms of its observed time-series depends only on the relative positions of the two terms, that is, on how far apart they are located from each other, and not on their absolute position~\cite{Chatfield}.}, whether $X$ causes $Y$ or not can be predicted by the log ratio of the prediction error variances:
\begin{equation}
    F_{X \rightarrow Y}=\ln \frac{\varepsilon}{\varepsilon_c}.
\end{equation}

This measure is called the F-statistic. If the model represented by equation~(\ref{eq_granger_causal_model}) is a better model for $Y(t)$ than equation~(\ref{eq_granger_noncausal_model}), then var($\epsilon_c$) $<$ var($\epsilon$) and $F_{X \rightarrow Y}$ will be greater than 0, suggesting that $X$ \emph{Granger causes} $Y$. Though this concept of causality uses an autoregressive model, due to the generic nature of the model making minimal assumptions about the underlying mechanism, this method is widely for data-driven causality estimation in diverse disciplines.

\subsection{Transfer Entropy (TE)}

Transfer Entropy quantifies the influence of process $J$ on transition probabilities of system $I$~\cite{Schreiber}. It measures the penalty to be paid in terms of excess amount of info-theoretic bits by assuming that the current state $i_{n+1}$ of a variable $I$ is independent of the past states $j_n^{(l)}$ of a variable $J$, i.e. assuming its distribution to be $q = p(i_{n+1}| i_{n}^{(k)})$ instead of $p(i_{n+1}| i_{n}^{(k)}, j_{n}^{(l)})$. Here $k$ and $l$ denote the number of past states of $I$ and $J$ respectively, on which the probability distribution of any state $i_{n+1}$ of process $I$ is dependent. Mathematically, 
\begin{equation}
TE_{J \rightarrow I}=\sum_{i,j}(p(i_{n+1},i_{n}^{(k)},j_{n}^{(l)})\log \frac{p(i_{n+1},i_{n}^{(k)},j_{n}^{(l)})}{p(i_{n+1},i_{n}^{(k)})}.
 \label{eq_TE}
\end{equation}
If $I$ and $J$ are independent processes, then $p(i_{n+1},i_{n}^{(k)},j_{n}^{(l)}) = p(i_{n+1},i_{n}^{(k)})$ for all $n, k, l$ and hence the above quantity will be zero. Intuitively, $TE_{J \rightarrow I}$ captures the flow of information from a process $J$ to a process $I$. In general, $TE_{J \rightarrow I} \neq TE_{I \rightarrow J}$.

\subsection{Convergent Cross Mapping (CCM)}

While GC has been developed for stochastic processes where the influences of different causal variables can be well separated, Convergent Cross Mapping is developed for deterministic processes that are not completely random. inspired from dynamical systems' theory, it can be applied even when causal variables have synergistic effects~\cite{Sugihara}. 

This method uses Takens' embedding theorem~\cite{Taken_Wiki} in a fundamental way. According to this theorem, observations from a single variable of the system can be used to reconstruct the attractor manifold of the entire dynamical system. CCM exploits the fact that two variables will be causally related if they are from the same dynamical system. If a variable $X$ causes $Y$, then the lagged (past) values of $Y$ can help to estimate states of $X$. This is true because of Taken's theorem -- manifold $M_Y$ (or $M_X$) of any one observed variable $Y$, will have a one to one mapping to the true manifold, $M$ and hence manifolds of two variables $M_Y$ and $M_X$ will have a one to one mapping to each other. However, this cross mapping is not symmetric. If $X$ is unidirectionally causing $Y$, past values of $Y$ will have information about $X$, but not the other way round. Thus, the state of $X$ will be predictable from $M_Y$, but $Y$ not from $M_X$.

\subsection{Compression Complexity Causality (CCC)}

\begin{figure}[!t]
\centering
\includegraphics[width=0.75\textwidth,trim={0 0 0 0},clip]{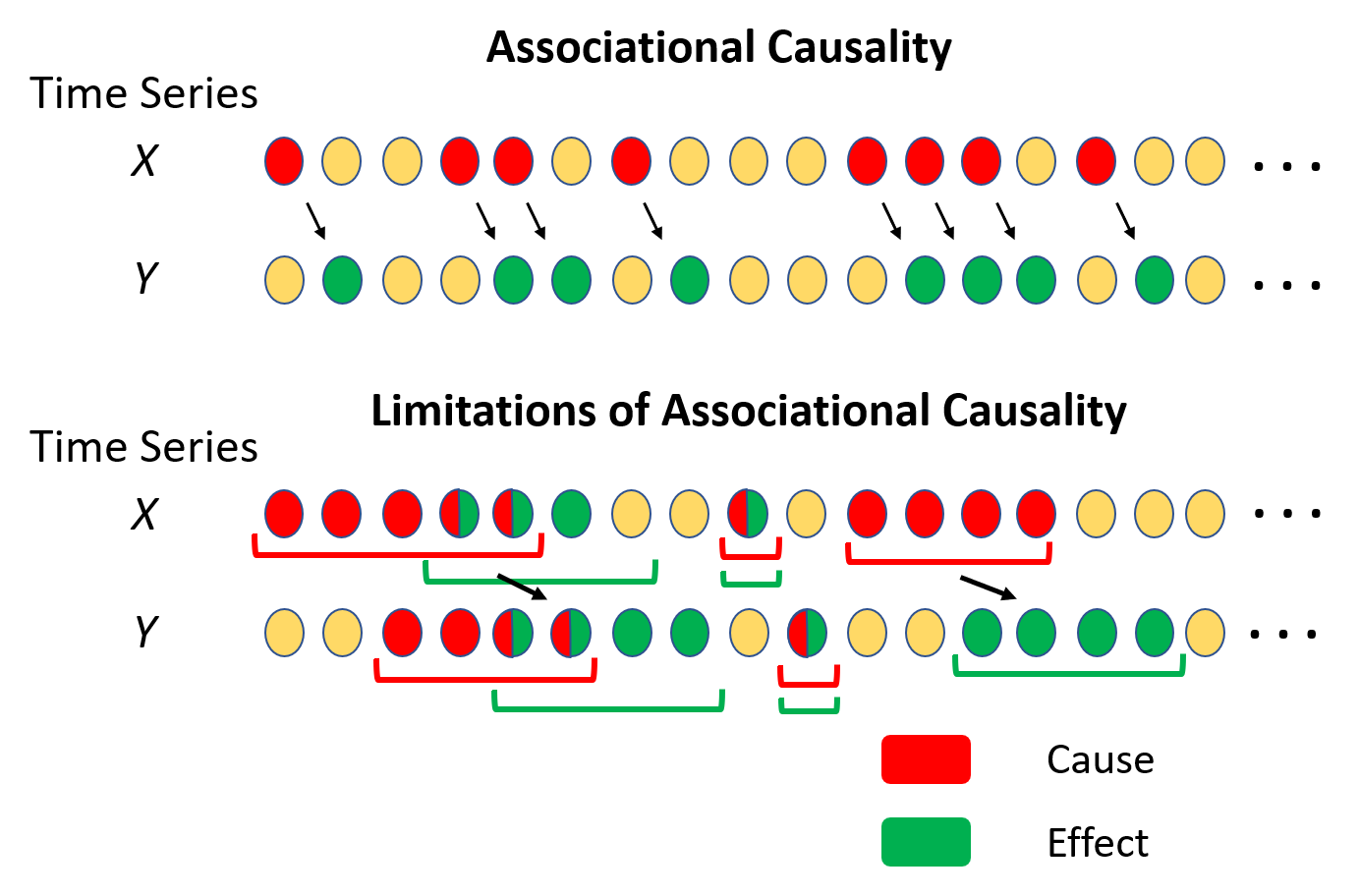}
\caption{Associational Causality is based on cause-effect separability of samples. It has limitations when cause-effect samples are overlapping. It may happen for continuous time processes which when sampled may result in  cause-effect information being simultaneously present in blocks of data or even in a single sample.}\label{fig_associational_causality}
\end{figure}

Measures such as GC and TE assume the inherent separability of `cause' and `effect' samples in time series data and are thus able to estimate only Associational Causality (Figure~\ref{fig_ladder}), which is at the first rung on the ladder of causation. However, many a times, cause and effect may co-exist in blocks of measurements or a single measurement. This may be the inherent nature of the dynamics of the process or a result of sampling being done at a scale different from the spatio-temporal scale of cause-effect dynamics (for example, during acquisition of measurements). Associational Causality measures are not appropriate in such scenarios. For a pair of time series $X$ and $Y$, with $X$ causing $Y$, an illustration of associatonal causality and its limitations are shown in Figure~\ref{fig_associational_causality}.

Compression-Complexity Causality (CCC), a recently proposed measure of causality does not make separability assumptions made by Associational Causality measures (such as GC and TE). In order to determine the causal influence of $X$ on $Y$, CCC captures how `dynamical complexities' of $Y$ change when information from the past of $X$ is brought in. CCC performs an intervention on $Y$, by inserting chunks of $X$ and stitching it with appropriate chunks of $Y$. This is the best that can be done with the data when it is not possible to intervene on the experimental set up~\cite{Kathpalia}. Thus CCC belongs to the second rung of the ladder of causation (Interventional Causality). 

In case of CCC, complexities of blocks of time series data are computed based on the measure \emph{Effort to Compress} (ETC)~\cite{Nagaraj}. To estimate CCC from $X$ to $Y$, we compute $CC(\Delta Y \vert Y_{past} )$ -- dynamical complexity of the current window of data, $\Delta Y$, from time series $Y$ conditioned with its own past, $Y_{past}$. This is compared with $CC(\Delta Y \vert Y_{past}, X_{past} )$ -- dynamical complexity of $\Delta Y$ conditioned jointly with the past of both $Y$ and $X$, $(Y_{past}, X_{past})$. Mathematically,
\begin{equation}
\label{eq_complexity_1D}
CC(\Delta Y \vert Y_{past} )=ETC(Y_{past}+\Delta Y)-ETC(Y_{past} ),
\end{equation}
\begin{equation}
\label{eq_complexity_2D}
CC(\Delta Y \vert Y_{past}, X_{past} ) = ETC(Y_{past}+\Delta Y, X_{past}+\Delta Y) - ETC(Y_{past},X_{past} ),
\end{equation}
\begin{equation}
\label{eq_ind_CCC}
CCC_{X_{past} \rightarrow \Delta Y}= CC(\Delta Y \vert Y_{past} ) - CC(\Delta Y \vert Y_{past}, X_{past} ).
\end{equation}
Averaged CCC from $Y$ to $X$ over the entire length of time series with the window $\Delta Y$ being slided by a step-size of $\delta$ is estimated as ---
\begin{equation}
\label{eq_CCC}
CCC_{X \rightarrow Y}=  \overline{CCC}_{X_{past} \rightarrow \Delta Y} = \overline{CC}(\Delta Y \vert Y_{past} ) - \overline{CC}(\Delta Y \vert Y_{past}, X_{past} ).
\end{equation}
\section{Model-based Causality Measures}

In cases where domain-knowledge is available and it is easy to perform lab-experiments to develop causal models underlying the generation of provided time series data, model-based causality estimation methods can be used. These kind of methods are both hypothesis (model) and data led and rest on comparison between assumed models and their optimization. Structural Equation Modeling (SEM)~\cite{Pearl1} and Dynamic Causal Modeling (DCM)~\cite{Friston} are examples of these kinds of methods. 

SEM includes a diverse set of computer algorithms, mathematical models and statistical methods that fit networks of constructs to data. The links between constructs of an SEM model may be estimated with independent regression equations or sometimes through more complicated techniques. In addition to being used in medicine, environmental science and engineering, SEM has also found applications in social science disciplines such as accounting and marketing. On the other hand, DCM was developed in the context of neuroscience. Its objective is to estimate coupling between different brain regions and to identify how the coupling is affected by environmental changes (i.e. say, temporal or contextual). Models of interaction between different cortical areas (nodes) are formulated in terms of ordinary or stochastic differential equations. The activity of the hidden states of these nodes maps to the measured activity based on another model. Bayesian model inversion is done to determine the best model and its parameters for the system using the acquired data.

\section{Conclusions and Way Ahead}
Since most studies in different disciplines are based on finding cause and effect relationships between the variables of the study, it is important for researchers to know how the means to extract causal relations have evolved over the years. In this study, we discuss different notions of causality, their order of hierarchy, estimation methods with a focus on determining causal relationships from time series data. We have discussed in detail data-driven, model-free methods of causality testing which are useful when data has been acquired and it is not possible to intervene on the experimental setup as well as when sufficient knowledge about the domain is unavailable. Model-based causality testing methods were also briefly discussed. These can be useful when along with time series data, background information of domain is also available.

While the discussed techniques are useful and are being used in widespread applications, it is important to employ them appropriately after carefully investigating whether the assumptions behind them hold for the given data. It is worthwhile to combine two or more notions or approaches of statistical causality estimation to obtain more reliable results as was done in the case of the famous study attributing smoking as a cause of lung cancer~\cite{Cornfield}. The ladder of causation by Judea Pearl is adapted for classification of time series causality testing methods. Intervention based approaches rank higher than Associational Causality measures. Other than model-based measures, CCC is the only data driven, model-free method which is included in this category. Going ahead, it will be useful to have more of such methods. Another future task would be combining the working of model-based and model-free methods of causality testing, which can be helpful to relax some of the assumptions made by model-based methods. The highest level in the ladder of causation is Counterfactual Causality, which involves imagination and creative thought experiments. This faculty, however, seems to be limited to humans and none of the time-series based causality methods to date are capable of accomplishing this.

Not only is the development and research of causality testing methods useful for immediate application to available real-world data, research in this is also important for more futuristic goals of the society, such as development of evolved and human-like artificial intelligence. Causal thinking is ingrained in our attitude and scientific research, but now the time has come that we also start thinking along the line of developing mathematical models and tools for \emph{retrocausality}, where the `future' can determine the `present'. 


\section*{Acknowledgement}
The authors gratefully acknowledge the financial support of ‘Cognitive Science Research Initiative’ (CSRI-DST) Grant No. DST/CSRI/2017/54(G) and Tata Trusts provided for this research. Aditi Kathpalia is thankful to Manipal Academy of Higher Education for permitting this research as part of the PhD programme.




\begin{thebibliography}{99} 
\bibitem{Cox_Wermuth_2004} 
David R. Cox, and Nanny Wermuth. 
\textit{Causality: A statistical view.} International Statistical Review 72(3): 285-305, 2004.
\bibitem{Good1}
Irving John Good. 
\textit{A causal calculus (I).} 
The British journal for the philosophy of science 11(44): 305-318, 1961.
\bibitem{Good2}
Irving John Good. 
\textit{A causal calculus (II).} 
The British journal for the philosophy of science 12: 43-51, 1962.
\bibitem{Suppes}
P. Suppes
\textit{A probabilistic theory of causation.}
Amsterdam: North Holland.
\bibitem{Granger}
 Clive WJ. Granger.
 \textit{Investigating causal relations by econometric models and cross-spectral methods.} 
 Econometrica: Journal of the Econometric Society: 424-438, 1969.
 \bibitem{Schweder}
 Tore Schweder. 
 \textit{Composable markov processes.} 
 Journal of applied probability 7(2): 400-410, 1970.
 \bibitem{Aalen}
 Odd O. Aalen. 
 \textit{Dynamic modelling and causality.} Scandinavian Actuarial Journal 1987, 3-4 :177-190, 1987.
 \bibitem{Fisher1}
 Ronald A. Fisher.
 \textit{The arrangement of field experiments.}
 Journal of the Ministry of Agriculture Great Britain, 33 :503-513, 1926.
 \bibitem{Fisher2}
 Ronald A. Fisher.
 \textit{Design of experiments.}
 Edinburgh: Oliver and Boyd. And subsequent editions, 1935.
 \bibitem{Yates1}
 Frank Yates. 
 \textit{Design and Analysis of Factorial Experiments.} 
 Harpenden: Imperial Bureau of Soil
Science, 1938.
\bibitem{Yates2}
 Frank Yates. 
\textit{Bases logiques de la planification des experiences.} 
Annals of the Institute of H. Poincar\'e,
12 :97–112, 1951.
\bibitem{Rubin}
Donald B. Rubin. 
\textit{Estimating causal effects of treatments in randomized and nonrandomized studies.} 
Journal of educational Psychology 66(5) :688, 1974.
\bibitem{Wright1}
Sewall Wright. 
\textit{Correlation and causation.} 
Journal of Agricultural Research, 20 :557–585, 1921.
\bibitem{Wright2}
Sewall Wright. 
\textit{The method of path coefficients.} Annals of Mathematical Statistics, 5 (3) :161–215, 1934.
\bibitem{Cochran}
William G. Cochran, and S. Paul Chambers. 
\textit{The planning of observational studies of human populations.} 
Journal of the Royal Statistical Society. Series A (General) 128(2) :234-266, 1965.
\bibitem{Pearl1}
Judea Pearl. 
\textit{Causality: models, reasoning and inference.}
29. Cambridge: MIT press, 2000.
\bibitem{Pearl2}
Judea Pearl, and Dana Mackenzie.
\textit{The book of why: the new science of cause and effect.}
Basic Books, 2018.
\bibitem{Schreiber}
Thomas Schreiber. 
\textit{Measuring information transfer.} Physical review letters 85(2) :461, 2000.
\bibitem{Kathpalia}
Aditi Kathpalia, and Nithin Nagaraj. \textit{Data-based intervention approach for Complexity-Causality measure.} 
PeerJ Computer Science 5 :e196, 2019.
\bibitem{Wiener}
Norbert Wiener.
\textit{The theory of prediction.}
Modern Mathematics for Engineers 1 :125-139, 1956.
\bibitem{Sugihara}
George Sugihara, Robert May, Hao Ye, Chih-hao Hsieh, Ethan Deyle, Michael Fogarty, and Stephan Munch. 
\textit{Detecting causality in complex ecosystems.} 
science 338(6106) :496-500, 2012.
\bibitem{Harnack}
Daniel Harnack, Erik Laminski, Maik Schünemann, and Klaus Richard Pawelzik. 
\textit{Topological causality in dynamical systems.} Physical review letters 119 (9) :098301, 2017.
\bibitem{Chatfield}
Chris Chatfield. 
\textit{The analysis of time series: an introduction.} 
Chapman and Hall/CRC, 2003.
\bibitem{Nagaraj}
Nithin Nagaraj, Karthi Balasubramanian, and Sutirth Dey. 
\textit{A new complexity measure for time series analysis and classification.} 
The European Physical Journal Special Topics 222, (3-4) :847-860, 2013.
\bibitem{Friston}
Karl J. Friston, Lee Harrison, and Will Penny. \textit{Dynamic causal modelling.} Neuroimage 19(4):1273-1302, 2003.
\bibitem{Taken_Wiki} 
Taken's Theorem, Wikipedia.
\\\texttt{https://en.wikipedia.org/wiki/Takens$\%$27s$\_$theorem}
\bibitem{Cornfield}
Jerome Cornfield, William Haenszel, E. Cuyler Hammond, Abraham M. Lilienfeld, Michael B. Shimkin, and Ernst L. Wynder. 
\textit{Smoking and lung cancer: recent evidence and a discussion of some questions.} Journal of the National Cancer institute 22(1) :173-203, 1959.
\end{thebibliography}
\end{document}